\documentclass[fleqn,twoside]{article}
\usepackage{espcrc2}

\usepackage{graphicx}

\newcommand{\AmS}{{\protect\the\textfont2
  A\kern-.1667em\lower.5ex\hbox{M}\kern-.125emS}}

\title{Generalized Parton Distributions and \\ the Spin Structure of the Nucleon}

\author{Xiangdong Ji\address{Department of Physics,
University of Maryland, College Park, Maryland 20742, USA}
	\thanks{Plenary talk given at Lattice 2002, Cambridge, MA, USA.
   This work is supported in part by funds provided by the
U.S.~Department of Energy (D.O.E.) under cooperative agreement
DOE-FG02-93ER-40762.}
	}
       
\begin{document}

\begin{abstract}
Generalized parton distributions are a type of hadronic observables
which has recently stimulated great interest among theorists and 
experimentalists alike. Introduced to delineate the spin structure
of the nucleon, the orbital angular momentum of quarks in particular,
the new distributions contain vast information about the internal structure
of the nucleon, with the usual electromagnetic form factors
and Feynman parton distributions as their special limits.
While new perturbative QCD processes, such as deeply virtual Compton 
scattering and exclusive meson production, have been found to measure
the distributions directly in experiments, lattice QCD offers a great
promise to provide the first-principle calculations of these 
interesting observables.
\vspace{1pc}
\end{abstract}

\maketitle

Since the EMC experiment on polarized deep-inelastic scattering (DIS)
on the proton in the late 1980s \cite{emc}, 
QCD spin physics has evolved into one of the most active 
frontiers in hadron physics. The International 
Spin Physics Conference is held every other year
with an average of 350 physicists participating in the meeting. The 
topics of the conference are often dominated by QCD spin physics. 
There are many experimental facilities around the world which have 
been playing a key role in spin physics research. For example, at CERN, 
following the EMC experiment there was SMC, and now the COMPASS experiment
is on its way to take data \cite{compass}. At SLAC, a series of experiments
were finished in the 1990s: E142, E143, E154, and E155. At 
DESY in Germany, the HERMES collaboration has taken data for 
a number of years, and new runnings have been planned (HERMES II).
A summary of the finished experiments can be found in \cite{filippone}. 
The spin physics at RHIC has just started, where 250 GeV polarized 
proton beams can be used to explore spin-dependent high-energy 
collisions \cite{saito}. Jefferson Lab in Virginia has a high intensity 6 GeV electron
beam, and many experiments there involve polarized beams and targets. 
A plan has been made to upgrade the beam energy to 12 GeV. MIT-Bates  
delivered a polarized electron beam a few years ago, and a number of 
intereting low-energy experiments have been done which have an impact
on spin physics.

There have been many important theoretical developments in spin
physics as well. Perturbative QCD analysis of $g_1(x)$ and $g_2(x)$ 
structure functions of the nucleon has been carried out to 
two- and one-loop orders, respectively \cite{g1,g2}. The role of 
the axial anomaly
in understanding the EMC results has been explored and debated 
\cite{anomaly}. A classification of the leading
and higher-twist polarized quark and gluon distributions has been
made \cite{classification}. QCD angular momentum and its role 
in high-energy scattering have been clarified \cite{angular}. 
The effects of parton transverse momentum in high-energy processes
have been explored systematically \cite{mulders}. Generalized parton 
distributions and deeply virtual Compton scattering 
are the main subject of this talk. 

One of the most important goals in QCD spin physics is to understand
the spin structure of the nucleon, i.e., how the nucleon spin is
made from its fundamental constituents. Because nonperturbative 
QCD tools, such as lattice QCD, is still under development, we have relied,
for many years, on model descriptions. In the naive SU(6) quark model, 
the three valence quarks
move in the $s$-orbit, and the spin of the nucleon comes entirely from the 
coupling of the quark spins. There are a number of experimental
evidences which seem to support this simple picture. For instance, the magnetic 
moment of the nucleon can be calculated in terms of the SU(6) wave 
function, and the ratio between the neutron and proton is \cite{moment}
\begin{equation}
          \mu_n/\mu_p = -2/3 \ . 
\end{equation}
The experimental data is $-0.68$, very close! Moreover, the $\Delta\rightarrow  N\gamma$
transition can take place with both M1 and E2 electromagnetic multipoles.
The quark model predicts that the E2 to M1 ratio is 0. Experimentally, 
the result is 1-3\%, depending on how the background contribution is 
modelled. [However, recent results from the large $N_c$ QCD seem to 
indicate that these quark model predictions follow simply from the contracted 
SU(4) symmetry present in the large $N_c$ limit, independent of the detailed
dynamics \cite{largen}]. 

This simple picture of the spin structure of the nucleon has been challenged 
by the polarized DIS data from CERN, SLAC and DESY
\cite{filippone}. In these experiments, the polarized quark distributions 
can be extracted:
\begin{equation}
    \Delta q(x) = q_+(x) - q_-(x) \ , 
\end{equation}
where $q_\pm(x)$ is the density of quarks with longitudinal momentum fraction 
$x$ and spin aligned (anti-aligned) to the spin of the nucleon. More precisely, 
the measured $g_1(x)$ structure function of the nucleon is related to the
helicity distributions by 
\begin{equation}
    g_1(x) = \frac{1}{2} \sum_f e_f^2 \Delta q_f(x) \ , 
\end{equation}
where $f$ sums over light quark flavors.
A summary of world data on the $g_1(x)$ structure functions
for proton, neutron and deuteron is shown in Fig. 1.

\begin{figure}[t]
\vspace{9pt}
\begin{center}
\includegraphics[width=6cm]{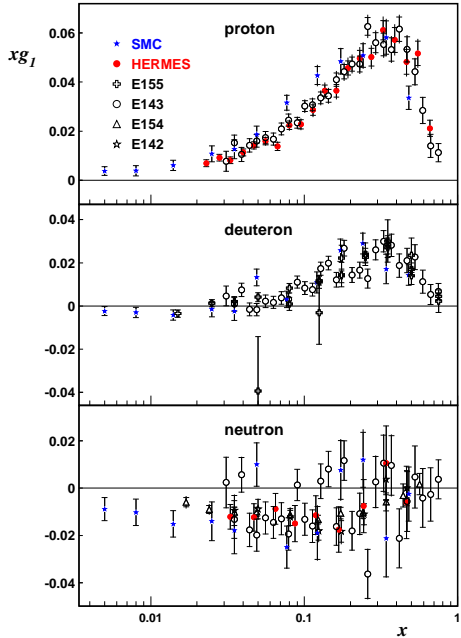}
\caption{A compilation of world data on the $g_1(x)$ structure function
of the proton, neutron ($^3$He), and deuteron.}
\end{center}
\label{fig:g1data}
\end{figure}

The operator production expansion (OPE) yields the deep-inelastic sum rule
\begin{equation}
       \int^1_0 g_1^p(x) dx = \frac{1}{18} (4\Delta u + \Delta d + \Delta s) \ , 
\end{equation}
where $\Delta q$ is the fraction of the nucleon spin carried by the quark spin
of flavor $q$. The total quark spin contribution to the proton spin
\begin{equation}
           \int^1_0 dx \sum_q \Delta q(x) = \Delta \Sigma = \Delta u + \Delta d + \Delta s \ , 
\end{equation}
cannot be extracted from the DIS data alone. One must use the axial couplings
$g_A$ determined from neutron and hyperon beta decay as well as the SU(3) flavor 
symmetry. An analysis of the world data to the next-to-leading order in perturbative
QCD yields \cite{filippone} 
\begin{equation}      
  \Delta \Sigma (1 ~{\rm GeV}^2) = 0.2 \pm 0.1 \ . 
\end{equation}
Therefore, at least 70\% of the proton spin resides in either the orbital motion
of the quarks or gluons.

To understand the spin structure of the nucleon 
in the framework of QCD, one must start from 
the QCD angular momentum operator in its gauge-invariant form \cite{ji1}:
\begin{equation}
    \vec{J}_{\rm QCD} = \vec{J}_{q} + \vec{J}_g \, ,
\end{equation}
where
\begin{eqnarray}
     \vec{J}_q &=& \int d^3x ~\vec{x} \times \vec{T}_q \nonumber \\
                 &=& \int d^3x ~\left[ \psi^\dagger
     {\vec{\Sigma}\over 2}\psi + \psi^\dagger \vec{x}\times
          (-i\vec{D})\psi\right]
     \, ,  \nonumber \\
     \vec{J}_g &=& \int d^3x ~\vec{x} \times (\vec{E} \times \vec{B}) \, .   
\label{ang}
\end{eqnarray}
The quark and gluon parts of the angular momentum
are generated from the quark and gluon
momentum densities $\vec {T}_q$ and $\vec{E}\times \vec{B}$,
respectively. $\vec{\Sigma}$ is
the Dirac spin-matrix and the corresponding term
is the quark spin contribution which we have discussed in the
context of the polarized DIS. $\vec{D}= \vec{\nabla}-
ig\vec{A}$ is the covariant derivative and the
associated term is the gauge-invariant quark
orbital contribution.

Using the above operator, one can easily construct
a decomposition for the spin of the nucleon. Consider
a nucleon moving in the $z$-direction, and 
in the helicity eigenstate $\lambda = 1/2$. The total
helicity can be calculated as an expectation value of
$J_z$ in the nucleon state,
\begin{equation}
        {1\over 2} = {1\over 2}\Delta \Sigma (\mu)
    + L_q(\mu) + J_q(\mu) \, ,     
\end{equation}
where the three terms denote the matrix elements
of three parts of the angular momentum operator
in Eq. (\ref{ang}). The physical significance
of each term is obvious, modulo the scale
and scheme dependence indicated by $\mu$. The scale
dependence in $\Sigma(\mu)$ is generated from
the U(1) axial anomaly \cite{anomaly}. Note that
the individual term in the above equation is
independent of the momentum of the nucleon.
In particular, it applies when the nucleon
is travelling with the speed of light (the infinite
momentum frame) \cite{ji2}.         

The DIS data $\Delta \Sigma<\!<1$ indicates that the nucleon 
is much more complicated than the naive quark model depicts. 
At a phenomenological level, one may 
regard the quark model as an effective description but it is of little 
use to explain the DIS data if one does not know the
relation between the QCD and constituent quarks. 
The challenge we have now is to find additional
contributions to the nucleon spin in the context 
of the above decomposition.

From the definition, we see that $J_{q,g}$ are just the 
matrix elements of the moment of the energy-momentum tensor
in a nucleon helicity state
\begin{equation}
      J_{q,g}(\mu) = \left\langle P{1\over 2} \left|
         \int d^3x (\vec{x}\times \vec{T}_{q,g})^z
 \right|P{1\over 2}\right\rangle \, ,
\label{matrix}
\end{equation}
I will argue below that the matrix elements can be extracted from the
form factors of the quark and gluon parts of
the QCD energy-momentum tensor $T^{\mu\nu}_{q,g}$
in the nucleon state.

To motivate the linkage, let's consider an analogous physics observable.
In the electromagnetism, the magnetic moment is defined as 
a spatial moment of the current density
\begin{equation}
 \vec{\mu} = \frac{1}{2} \int d^3x \vec{x} \times \vec{j} \ ,
\end{equation}
If one knows the Dirac and Pauli     
form factors of the electromagnetic current,
$F_1(Q^2)$ and $F_2(Q^2)$,
\begin{eqnarray}
      \langle P'|j^\mu|P\rangle
  &=& \overline{U}(P') \left[ F_1(t)\gamma^\mu \right. \nonumber \\
      &&\left.     + F_2(t) \frac{i\sigma^{\mu\nu}\Delta_\nu}{2M}
            \right]U(P) \ , 
\end{eqnarray}
where $t=(P-P')^2$, and $\Delta= P-P'$. 
The spin-flip form factor $\Delta G_M(t)=q(F_1(t) + F_2(t))$ 
yields the electric current distribution in the nucleon. 
From this, we can calculate the magnetic moment
\begin{equation}
         \mu_p = G_M(0) = F_1(0) +F_2(0)\ ,  
\end{equation}
in units of $e\hbar/2Mc$.

Using Lorentz and discrete symmetries, it is easy to show that
the symmetric and traceless part of the quark and gluon energy 
momentum tensors each supports three form factors,
\begin{eqnarray}
  \langle P'| T_{q,g}^{\mu\nu} |P\rangle
   &=& \overline U(P') \Big[ A_{q,g}(t) \gamma^{(\mu} \overline P^{\nu)}
     \nonumber \\ &&  +
  B_{q,g}(t) \overline P^{(\mu} i\sigma^{\nu)\alpha}\Delta_\alpha/2M \nonumber \\
 && + C_{q,g}(t)\Delta^{(\mu} \Delta^{\nu)}/M \Big] U(P) \, , 
\end{eqnarray}
where $\overline{P}=(P+P')/2$. Taking the forward limit of the $\mu=0$ 
component and integrating over 3-space, one finds that $A_{q,g}(0)$ give   
the momentum fractions of the nucleon carried by
quarks and gluons (the momentum sum rule contraints $A_q(0)+A_g(0)= 1$).
On the other hand, substituting
the above into the nucleon matrix element in Eq. (\ref{matrix}),
one finds \cite{ji1}
\begin{eqnarray}
      J_{q, g}(\mu^2) = {1\over 2} \left[A_{q,g}(0,\mu^2) + B_{q,g}(0,\mu^2)\right] \, .
\end{eqnarray}
Therefore, the matrix elements of the energy-momentum
tensor yield the fractions of
the nucleon spin carried by quarks and gluons.

While a measurment of the gravitational form factors is
impossible in the forseeable future, they can be extrated from
QCD hard scattering. A process which was first identified
is deeply virtual Compton scattering \cite{ji1}, in which 
a spacelike virtual photon created through lepton scattering,
scatters off the nucleon target, yielding a final state of a real
photon and a recoiled proton (see Fig. 2). 
When the virtual photon is in the Bjorken limit, the scattering 
process simplifies enormously: the single quark scattering 
dominates any other subprocesses
and we have essentially a Compton scattering on a single quark. 
Compared with DIS in which the nucleon breaks into many fragments, 
DVCS in a sense is a non-invasive surgery. 

\begin{figure}[t]
\vspace{9pt}
\begin{center}
\includegraphics[width=7cm]{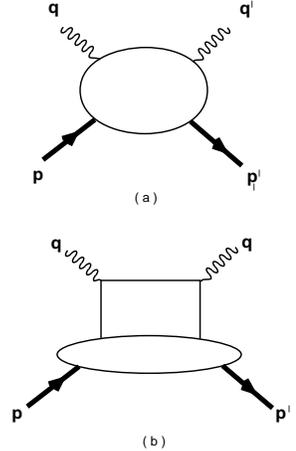}
\caption{(a) Compton scattering on a nucleon; (b) Leading hand-bag contribution
to DVCS.}
\end{center}
\label{fig:dvcs}
\end{figure}

The relation of the DVCS amplitude to the energy-momentum form factors can be 
seen as follows. The Compton scattering amplitude is a matrix element
of the time-ordered product of the electromagnetic
currents,
\begin{equation}
    T^{\mu\nu} = i\int d^4z e^{\bar q\cdot z}
    \left\langle P'\left|{\rm T}J^\mu\left(-{z\over 2}\right)
    J^\nu\left({z\over 2}\right)\right|P\right\rangle
\end{equation}
where $\overline q = (q+q')/2$. In the Bjorken
limit, $-q^2 $ and $ P\cdot q\rightarrow
\infty$ and their ratio remains finite,
we can use the operator product expansion to 
approximate the current product, 
\begin{equation}
        TJ^\mu(\xi) J^\nu(0) = ... + C^{\mu\nu\alpha\beta}(\xi^2)
               \overline{\psi}\gamma_\alpha iD_\beta \psi + ....
\end{equation}
where $C^{\mu\nu\alpha\beta}$ is the coefficient function and other 
local operators have been omitted. Therefore, the DVCS amplitude
contains the matrix element of the quark 
energy-momentum tensor \cite{ji6}. The gluon contribution 
comes at the next-to-leading order. However, as the ellipses 
indicate, the DVCS amplitude contains much more information than 
just the energy-momentum tensor. It contains a whole new
type of parton distributions. 

To motivate the distributions, let us consider a form factor of a 
local current $\overline{\psi}\Gamma \psi(0)$. Its matrix element
between the nucleon states $|P\rangle$ and $|P'\rangle$ can be 
obtained (at least in the infinite momentum frame for a good component
of the current) by annihilating
a quark of momentum $\vec{k}$ at the spacetime point 0 and creating 
the same particle back at the same point with a new momentum $\vec{q}+\vec{k}$. 
On the other hand, the ordinary Feynman 
parton distribution $q(x)$ can be interpreted as a forward matrix element 
in the nucleon state in which a quark of momentum $xp^\mu$ is annihilated
at point 0 and created back with the same momentum at a different spacetime 
point $\lambda n^\mu$, where $n$ is a light-cone vector $n^2=0$ conjugating
to the nucleon momentum $p$.

The generalized parton distributions (GPDs) combine the spacetime structures 
of the above matrix elements. They are defined as an interference amplitude 
in which a quark is annihilated at spacetime point 0 with momentum $xp^\mu$, 
and created back at another point $\lambda n^\mu$ with a new momentum $xp^\mu+q^\mu$, 
yielding a recoiled nucleon. Therefore, the
GPDs naturally include the elastic form factors and Feynman parton 
distributions in their kinematic limits. In fact, they contain
much more information than the two traditional types of observables.

Technically, GPDs can be defined through the matrix elements of a 
bilocal light-cone operator \cite{ji1,muller}, 
\begin{eqnarray}
   && {1\over 2}\int {d\lambda\over 2\pi} 
e^{i\lambda x} 
\left\langle P'\left|\overline \psi_q \left(-{\lambda \over
2}n\right)
      \not \! n  \right. \right.\nonumber \\
    && \times 
    \left.  \left.  {\cal P}e^{-ig\int^{-\lambda/2}_{\lambda/ 2}
       d\alpha ~n\cdot A(\alpha n)}
    \psi_q\left({\lambda \over 2}n\right) \right| P\right\rangle
    \nonumber \\
  && = H_q(x, \xi, t)~ {1\over 2}\overline U(P')\not\! n U(P)
    + E_q(x, \xi, t)~ \nonumber \\
&& ~~~ \times {1\over 2}\overline U(P') {i\sigma^{\mu\nu}
  n_\mu \Delta_\nu \over 2M} U(P) \, .
\label{string}
\end{eqnarray}
The light-cone bilocal operator (or light-ray
operator) arises frequently in hard scattering processes
in which partons propagate along the light-cone. 
The parton distributions are most naturally defined
in terms of the matrix elements of the bilocal operator.
In this context, the Feynman $x$ is just the conjugating
Fourier variable of the light-cone distance. The Lorentz structures
in the second line in the above equation are independent
and complete.      

The GPD's are more complicated than the
Feynman parton distributions because of their
dependence on the momentum transfer $\Delta$. As such, they 
contain two more scalar variables besides
the Feynman variable $x$. The variable $t$ is the usual $t$-channel
invariant which is always present in a form factor.
The $\xi$ variable is a natural product of marrying the concepts
of the Feynman distribution and form factor: The former requires
the presence of a preferred momentum $p^\mu$ along which the partons are
predominantly moving, and the latter requires a four-momentum
transfer $\Delta$; $\xi$ is just a scalar product of the two
momenta. 

Since the quark and gluon energy-momentum tensors
are just the twist-two, spin-two, parton helicity-independent
operators, we immediately have the following
sum rule from the off-forward distributions;
\begin{eqnarray}
    && \int^1_{-1} dx x [H_q(x, \xi, t) +
       E_q(x, \xi, t) ] \nonumber \\
    & =& A_q(t) + B_q(t) \, ,
\end{eqnarray}
where the $\xi$ dependence, or $C_q(t)$
contamination, drops out. Extrapolating the sum rule
to $t=0$, the total quark (and hence quark orbital)
contribution to the nucleon spin is obtained.
A similar sum rule exists for gluons.    
Thus a deep understanding of the spin structure of the
nucleon may be achieved by measuring the GPDs
from high energy experiments.                      
               
Recently, M. Burkardt has constructed an interpretation
of GPD in the coordinate space \cite{burkardt}. 
Consider the nucleon state localized
in the transverse plane at $r_\perp=0$, 
\begin{equation}
     |p^+, r_\perp=0 \rangle
   = N\int d^2p_\perp |p^+,p_\perp\rangle \ , 
\end{equation}
where $N$ is a normalization factor.
Define a parton distribution $q(x, b_\perp)$ which is the density
of partons with longitudinal momentum $xp^+$ and the transverse distance
$b_\perp$ in the state,
\begin{eqnarray}
  &&  q(x, b_\perp)
   = \int {d\lambda \over 4\pi}
 e^{ix\lambda} \nonumber \\
 && ~~~ \times \langle p^+, r_\perp|\overline{\psi}(0, b)
   \gamma^+ \psi(\lambda n, b)|p^+,r_\perp\rangle. 
\end{eqnarray}
Then it can be shown that $q(x, b)$ is the Fourier transformation of $H(x,-\Delta_\perp^2,
\xi=0)$ with resprect to the transverse momentum transfer, 
\begin{equation}
     q(x, b) = \int d^2\Delta_\perp
            H(x, -\Delta^2_\perp) e^{ib\cdot \Delta_\perp} \ . 
\end{equation}
Thus the GPDs provide the transverse location of the partons in
the nucleon.
    
From the viewpoint of the low-energy nucleon structure, 
it is, perhaps, most interesting to consider
GPDs as the generating functions for the form factors of 
the so-called twist-two operators. 
Recall that the matrix elements of the electromagnetic
current in the same nucleon state are 
determined by symmetry, whereas those in the unequal momentum
states define the (Dirac and Pauli) form factors which
contain such interesting information as the charge radius
and magnetic moment of the nucleon. The following
tower of twist-two operators represents a generalization 
of the electromagnetic current
\begin{equation}
    {\cal O}^{\mu_1\cdots\mu_n}_q =  
       \overline \psi_q i\stackrel{\leftrightarrow}{\cal D}^{(\mu_1}
        \cdots  i\stackrel{\leftrightarrow}{\cal D}^{\mu_{n-1}}
        \gamma^{\mu_n)} \psi_q\ ,
\label{O1} 
\end{equation}
where all indices are symmetrized and traceless (indicated
by $(\cdots)$) and $\stackrel{\leftrightarrow}{\cal D}
= (\stackrel{\rightarrow}{\cal D} - \stackrel{\leftarrow}{\cal D})/2. $
Technically, these operators transform as $(n/2,n/2)$ 
of the Lorentz group. They appear, as we have discussed
before, in the 
operator production expansion of the two electromangetic 
currents. Thus, although these generalized 
currents do not couple directly to any known fundamental 
interactions, they can nonetheless be studied 
indirectly in hard scattering processes.

Since the operators for $n>1$ 
are not related to any symmetry in the QCD
lagrangian, their matrix elements between 
the equal momentum states, 
\begin{equation}
  \langle  P | {\cal O}^{\mu_1\cdots\mu_n}_q|P\rangle =  
      2a_n(\mu) P^{(\mu_1} \cdots P^{\mu_n)}\ , 
\end{equation}
contain valuable dynamical information about the
internal structure of the nucleon. The $\mu$ 
dependence of the above matrix elements signifies
the dependence on renormalization scales and schemes. 
The quark distribution $q(x, \mu)$ introduced 
by Feynman has a simple connection to the 
above matrix elements: 
\begin{equation}
       \int^1_{-1} dx x^{n-1} q(x,\mu) = a_n(\mu) \ ,   
\end{equation}
where $q(x,\mu)$ is chosen to have support 
in $(-1,1)$. For $x>0$, $q(x, \mu)$ is the 
density of quarks which carry the $x$ fraction of 
the parent nucleon momentum. The density of antiquarks 
is customarily denoted as $\bar q(x,\mu)$, which in 
the above notation is $-q(-x,\mu)$. 

Just like the form factors of the electromagnetic
current, additional information about the nucleon structure 
can be found in the form factors of the twist-two operators
when the matrix elements are taken between the states of
unequal momenta. Using Lorentz symmetry and parity and
time reversal invariance, one can write down all possible
form factors of the spin-$n$ operator \cite{melnithouk,ji:1998pc,lebed}
\begin{eqnarray}
&&\langle P'| O^{\mu_1\cdots \mu_n}_q |P\rangle 
  \nonumber \\
   && = {\overline U}(P') \gamma^{(\mu_1} U(P) 
     \nonumber \\ &&
 \times  \sum_{i=0}^{[{n-1\over 2}]}
       A_{qn,2i}(t) \Delta^{\mu_2}\cdots \Delta^{\mu_{2i+1}} 
      \overline{P}^{\mu_{2i+2}}\cdots
      \overline{P}^{\mu_n)}  \nonumber \\ 
    &&  + ~  {\overline U}(P'){\sigma^{(\mu_1\alpha} 
     i\Delta_\alpha \over 2M}U(P)  \nonumber \\ &&
       \times \sum_{i=0}^{[{n-1\over 2}]}
     B_{qn,2i}(t) \Delta^{\mu_2}\cdots \Delta^{\mu_{2i+1}} 
      \overline{P}^{\mu_{2i+2}}\cdots
    \overline{P}^{\mu_n)} \nonumber \\
     &&  + ~ C_{qn}(t) ~{\rm Mod}(n+1,2)~{1\over M}\bar U(P') U(P) 
  \nonumber \\ && \times \Delta^{(\mu_1} \cdots \Delta^{\mu_n)} \ , 
\label{form}
\end{eqnarray}
where ${\overline U}(P')$ and $U(P)$ are the Dirac spinors
and ${\rm Mod}(n+1,2)$ is 1 when $n$ even, 0 when $n$ odd.
For $n\ge 1$, even or odd, there are $n+1$ 
form factors. $C_{qn}(t)$ is present only when $n$ is even.

Like the forward matrix elements $a_n$, the above form 
factors define GPDs completely. 
Introduce a light-like vector $n^\mu$, 
which is conjugate to $\overline P$ in the sense
that $\bar P \cdot n = 1$. Write $\overline P = 
p + (\overline M^2 /2) n $, where $\overline M^2 = M^2-t/4$
and $p$ is another light-like vector.  
Contracting both sides of Eq. (\ref{form}) with $n_{\mu_1}
\cdots n_{\mu_n}$, we have
\begin{eqnarray}
&& n_{\mu_1}\cdots n_{\mu_n} \langle P'| O^{\mu_1\cdots \mu_n}_q |P\rangle
   \nonumber \\
  && =   {\overline U}(P') \not\!n U(P)   
     H_{qn}(\xi, t) \nonumber \\
 &&
  + {\overline U}(P'){\sigma^{\mu\alpha} n_\mu i\Delta_\alpha 
    \over 2M}U(P) E_{qn}(\xi, t) \ , 
\end{eqnarray}
where 
\begin{eqnarray}
   H_{qn} (\xi,t) &=& \sum_{i=0}^{[{n-1\over 2}]}
     A_{qn,2i}(t) (-2\xi)^{2i} \nonumber \\
 &&+   {\rm Mod}(n+1,2)~ C_{qn}(t) (-2\xi)^n \ , \nonumber \\
   E_{qn} (\xi,t) &=& \sum_{i=0}^{[{n-1\over 2}]}
     B_{qn,2i}(t) (-2\xi)^{2i} \nonumber \\ && 
    - {\rm Mod}(n+1,2)~ C_{qn}(t) (-2\xi)^n \ .  
\end{eqnarray}
Here we have defined a new variable $\xi = -n\cdot 
\Delta/2$. Clearly, the $\xi$
dependence of $H_{qn}$ and $E_{qn}$ helps to 
distinguish among the different form factors of the same operator.
Now the moments of $H(x, \xi, t)$ and $E(x, \xi,t)$ are simply
\begin{eqnarray}
     \int^1_{-1} dx x^{n-1} E_q(x,\xi,t) & = &E_{qn}(\xi, t) \ , \nonumber \\
    \int^1_{-1} dx x^{n-1} H_q(x,\xi,t) & = &H_{qn}(\xi, t) \ . 
\end{eqnarray}     
Since all form factors are real, the new distributions
are consequently real. Moreover, because of time-reversal 
and hermiticity, they are {\it even} 
functions of $\xi$.

The Compton amplitude can be expressed in terms of the GPDs. 
In the leading order, 
\begin{eqnarray}
&&T^{\mu\nu} = g^{\mu\nu}_\perp \int^1_{-1}
       dx \left({1\over x-\xi+ i\epsilon}
       + {1\over x+\xi-i\epsilon}\right)
\nonumber \\
  && \times  \sum_q
       e_q^2 F_q(x, \xi, t, Q^2)
    \nonumber \\
 &&  +  i \epsilon^{\mu\nu\alpha\beta} p_\alpha n_\beta
      \int^1_{-1}
      dx  \left({1\over x-\xi + i\epsilon}
       - {1\over x+\xi-i\epsilon}\right)  \nonumber \\
&& \times  \sum_q e_q^2
      \tilde F_q(x, \xi, t, Q^2) \, ,
\end{eqnarray}
where $n$ and $p$ are the conjugate light-cone vectors
defined according to the collinear direction of
$\overline P$, and $g^{\mu\nu}_\perp$
is the metric tensor in the transverse space. $\xi$ is related   
to the Bjorken variable $x_B = - q^2 /(2 P\cdot q)$
by $x_B=2\xi/(1+\xi)$. In addition to DVCS, the electromagnetic 
radiation from the scattering lepton, the so-called Bethe-Heitler process, also produces
the same final state (background). 

The first evidence for 
DVCS was seen by H1 and Zeus collaborations at the 
HERA collider, DESY \cite{saull,adloff}. In Fig. 3, we have 
shown the differential
cross section for $e^+p\rightarrow e^+\gamma p$
as a function of $Q^2$ from H1. The hatched histogram shows 
the contribution of the Bethe-Heitler process. 
There is a clear access of events above the background.

\begin{figure}[thb]
\vspace{9pt}
\begin{center}
\includegraphics[width=6cm]{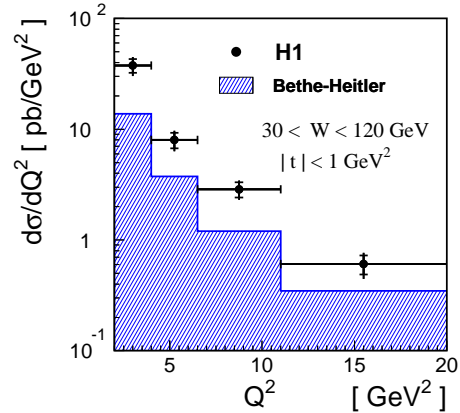}
\caption{Cross section measurements for the $\gamma^*p\rightarrow
\gamma p$ DVCS process as a function of $Q^2$ \cite{adloff}.}
\end{center}
\label{fig:h1}
\end{figure}

\begin{figure}[thb]
\vspace{9pt}
\begin{center}
\includegraphics[width=6cm]{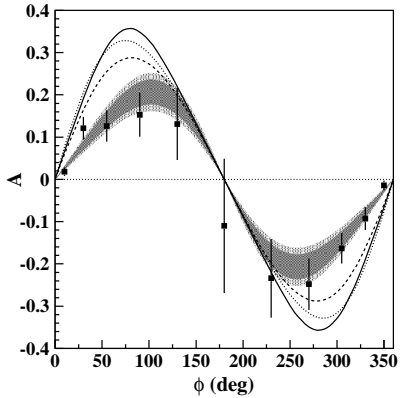}
\caption{Single spin asymmetries measured by the CLAS collaboration
which is sensitive to the inteference between Bethe-Heitler and DVCS processes.}
\end{center}
\label{fig:clas}
\end{figure}

At low-energy facilities such as JLab, the Bether-Heitler
process overwhelms the cross section. In this case, 
one can isolate the interference contribution from the Bethe-Heitler 
and DVCS processes. Shown in Fig. 4 is the single beam spin asymmetry
measured by CLAS as a function of the angle between the 
virtual and outgoing photons \cite{stepanyan}. A similar data exist from the
HERMES collaboration \cite{airapetian}. More
experiments on DVCS will be done in the future at HERA,
CERN, and Jefferson Lab. 

An all-order proof of the DVCS factorization was discussed in \cite{rad2}
and in \cite{ji6,col4}.
Factorization concerns separation of soft physics at the scale of hadron
masses and perturbative physics at the scale of probe in the DVCS
amplitude. This can be done at the same level as for inclusive 
DIS. Therefore experimental cross section can be used to directly extract
GPDs. This is nontrivial beause the final photon here is on-shell.
Technically, it amounts to showing all soft physics either can be absorbed
into the GPDs, which are nonperturbative anyway, or are down by powers of 
$1/Q^2$. The factorization result can be formulated in terms of an 
operator product expansion familiar in the field theory textbooks.

The next-to-leading order corrections to DVCS are
important for a precision extraction of GPDs and for an 
estimate of scaling violation. The complete result is now known.  
The one-loop corrections to DVCS have
first been studied by Osborne and Ji \cite{ji6}, and also by 
Belitsky and Mueller \cite{bm1}. 
Two-loop anomalous dimensions were obtained 
by Belitsky and Muller \cite{bm2}. It is the same as that
for inclusive DIS except for the non-diagonal contribution 
which can be determined by the one-loop conformal anomaly. 
Classical chromodynamics is invariant
under comformal symmetry which is broken by quantum mechanical effects. 
Because of this, an operator can acquire a scaling dimension under the
spatial conformal transformation (conformal anomaly). The size for
the NLO corrections has been estimated in \cite{bm3}. 

The leading higher-twist contribution to DVCS comes from twist-three longitudinal
photon scattering. In the interference amplitude, the imaginary part has a 
characteristic $\sin 2\phi$ and real part $\cos 2\phi$ dependence. The contribution
can be estimated in the so-called Wilczek-Wendzura approximation (neglecting
dynamical higher-twist effects). Belitsky, Mueller, and Kirchner et al.  \cite{bm4}
found that the correction to the single spin asymmetry from the beam polarization
is about 6\%, and from the target polarization the correction is about 9\%. For the spin-averaged
cross section it is 17\%, and 3\% for the double spin asymmetry.                                                                                     

DVCS can be generalized to the case of exclusive production
of mesons \cite{rad1}. Collins, Frankfurt and Strikman have shown that the 
deep-exclusive meson production is factorizable to all orders
in perturbation theory \cite{cfs}. The vector meson production entails a 
rich spin and flavor structure. For example, the vector meson 
production is sensitive to quark helicity-independent distributions, 
whereas the pseudo-scalar mesons are sensitive to quark helicity-dependent 
distributions. The next-to-leading order perturbative 
QCD correction for pseudo-scalar meson production has first been calculated
by Belitsky and Mueller \cite{bm4}.
Meson production may be easier to detect; however,
it has a twist suppression, $1/Q^2$.
In addition, the theoretical cross section
depends on the unknown light-cone meson wave function. 
Recently, it has been found that a quantity very sensitive to 
the quark angular momentum is the target transverse spin asymmetry for 
vector meson production \cite{vander}.The
asymmetry comes from the interference of two DVCS amplitudes and
is linear in the $E(x,\xi, t)$ distribution. Since the asymmetry involves
the ratio, it is insensitive to the next-to-leading order 
and higher-twist effects. The vector meson final state allows 
separation of different quark flavors. For example, the
$\rho^0$ production is sensitive to the combination $2J_u + J_d$, 
the $\omega$ meson is sensitive to $2J_u-J_d$, and finally 
the $\rho^+$ meson is sensitive to the $J_u-J_d$ combination.

To summarize, the recently discovered generalized parton
distributions are a new type of nucleon observables which
contain rich information about the internal structure of the
nucleon. Many experiments have been planned to measure the distributions. 
Lattice QCD will be a unique theoretical tool to 
calculate the distribution from the first principles \cite{lattice}.

\end{document}